\documentclass[conference]{IEEEtran}
\IEEEoverridecommandlockouts
\usepackage{cite}
\usepackage{amsmath,amssymb,amsfonts}
\usepackage{algorithmic}
\usepackage[utf8]{inputenc}
\usepackage{graphicx}
\usepackage{textcomp}
\usepackage{xcolor}
\usepackage{url}
\usepackage{amsthm}

\def\BibTeX{{\rm B\kern-.05em{\sc i\kern-.025em b}\kern-.08em
    T\kern-.1667em\lower.7ex\hbox{E}\kern-.125emX}}

\IEEEoverridecommandlockouts
\IEEEpubid{\makebox[\columnwidth]{978-1-7281-2522-0/19/\$31.00~\copyright2019 IEEE \hfill} \hspace{\columnsep}\makebox[\columnwidth]{ }}

\begin{document}

\title{Anomalous Communications Detection in IoT Networks Using Sparse Autoencoders
}

\author{Mustafizur R. Shahid\IEEEauthorrefmark{1}, Gregory Blanc\IEEEauthorrefmark{1}, Zonghua Zhang\IEEEauthorrefmark{2}, Hervé Debar\IEEEauthorrefmark{1}\\
\IEEEauthorrefmark{1}SAMOVAR, CNRS, Télécom SudParis, Institut Polytechnique de Paris, France\\
\{mustafizur.shahid, gregory.blanc, herve.debar\}@telecom-sudparis.eu \\ 
\IEEEauthorrefmark{2}SAMOVAR, CNRS, IMT Lille-Douai, Institut Mines-Télécom, France\\
zonghua.zhang@imt-lille-douai.fr\\
}




\newcommand{\boxedtext}[1]{\fbox{\scriptsize\bfseries\textsf{#1}}}
\newcommand{\myremark}[2]{
   \textcolor{blue}{\boxedtext{#1}
      {\small$\blacktriangleright$\emph{\textsl{#2}}$\blacktriangleleft$}
}}
\newcommand{\DELETE}[2]{
   \textcolor{red}{
         {\small$\blacktriangleright$\st{#2}$\blacktriangleleft$}
    }
}
\newcommand\GB[1]{\myremark{GB}{#1}}

\maketitle

\begin{abstract}
Nowadays, IoT devices have been widely deployed for enabling various smart services, such as, smart home or e-healthcare. However, security remains as one of the paramount concern as many IoT devices are vulnerable. Moreover, IoT malware are constantly evolving and getting more sophisticated. IoT devices are intended to perform very specific tasks, so their networking behavior is expected to be reasonably stable and predictable. Any significant behavioral deviation from the normal patterns would indicate anomalous events. In this paper, we present a method to detect anomalous network communications in IoT networks using a set of sparse autoencoders. The proposed approach allows us to differentiate malicious communications from legitimate ones. So that, if a device is compromised only malicious communications can be dropped while the service provided by the device is not totally interrupted. To characterize network behavior, bidirectional TCP flows are extracted and described using statistics on the size of the first N packets sent and received, along with statistics on the corresponding inter-arrival times between packets. A set of sparse autoencoders is then trained to learn the profile of the legitimate communications generated by an experimental smart home network. Depending on the value of N, the developed model achieves attack detection rates ranging from 86.9\% to 91.2\%, and false positive rates ranging from 0.1\% to 0.5\%.
\end{abstract}


\section{Introduction}
The total number of IoT devices is expected to reach 75 billion by 2030 \cite{IoTforecast}. This rapid adoption of IoT introduces new security challenges \cite{ZhangIoTSecurityChallengesOppor}\cite{KHANIoTSecurityReviewBlockchainChallenges} for network administrators. Most IoT devices are vulnerable because of a lack of security experience of the manufacturer and a short time to market. The devices present many security flaws, such as weak passwords, backdoors and various software vulnerabilities \cite{wangInsideLookIoTMalware}\cite{owaspIoT}\cite{bertinoBotnetsIoTSecurity}. 
Most vulnerabilities are not even properly patched because of a poor software update policy or a lack of security awareness of the end user. 
From 2016 to 2017, there was a 600 percent increase in IoT attacks \cite{symantecISTR}. In 2016, Mirai turned thousands of smart devices into bots. Those infected devices were primarily used to perform DDoS attacks \cite{AntonakakisMiraiBotnet}. Mirai performed brute-forcing to infect poorly configured devices. Since Mirai, IoT malware have considerably evolved and are getting more and more complex. In 2017, the Reaper IoT botnet was mainly exploiting disclosed vulnerabilities to infect devices \cite{ReaperBotnet}. In 2018, HideNSeek has become the first IoT malware capable of surviving a device reboot \cite{HideNSeekBotnet}. To cope with these permanently evolving threats, new anomaly detection methods need to be developed, especially designed to deal with the huge diversity of IoT devices. 

In this context, machine learning can be leveraged to develop techniques to detect malicious activities in IoT networks. An IoT device is intended to perform very specific tasks that remain the same over time. For example, a smart bulb can only be switched on or off. However, it is not supposed to send emails or click on ads, making its behavior very stable and predictable.
On the contrary, the main difficulty encountered in applying data analysis techniques for intrusion detection in the case of general purpose devices, such as desktop computers, laptops or smartphones, is the great variability and diversity of the generated network traffic \cite{OutsideTheClosedWorld}. The versatile nature of the network traffic generated by general purpose devices makes it difficult to define what behavior is supposed to be legitimate. 
Therefore, anomaly detection methods are, thus, better suited for IoT devices. 

In this paper, we present a method to detect anomalous communications in IoT networks. Communications detected as being anomalous can be dropped while the legitimate communications are maintained. Such a model, that identifies and blocks only anomalous connections, will allow a device to provide services even if it is infected by a malware. We propose to take advantage of a set of sparse autoencoders to learn the legitimate communication behavior of the IoT network. Due to the great diversity of IoT devices (camera, smart bulb, motion sensor, etc), the network communications behavior can vary greatly from one device to another. Therefore, one different sparse autoencoder is trained for each IoT device type. However, during the testing phase (or actual deployment phase), we assume that it might not be possible to determine what device the network communications belong to. Hence, we feed the network communications data to all the trained sparse autoencoders. A communication is considered to be anomalous if all the trained sparse autoencoders consider so. We set up an experimental smart home network composed of four IoT devices to generate legitimate network traffic data. Bidirectional TCP flows are then extracted. Features used to describe the bidirectional flows are statistics on the size of the first N packets sent and received, as well as, statistics on the inter-arrival times between those packets. Depending on the value of N, the developed model achieves an attack detection rate between 86.9\% and 91.2\%, and a false positive rate between 0.1\% and 0.5\%. 

The rest of this paper is organized as follows: Section \ref{AE AD} describes how to use sparse autoencoders to perform anomaly detection. In Section \ref{Proposed Model}, we describe our model and the features that we use. Section \ref{dataset description} describes the data used to train and test the model. Then, in Section \ref{experimental results} we present the experimental results. Section \ref{Related Work} reviews the related works. Finally, Section \ref{Conclusion and Future Works} concludes and presents the possible future works.

\section{Sparse Autoencoders for Anomaly Detection} \label{AE AD}

\subsection{Sparse Autoencoder} \label{training autoencoder}

\begin{figure}[]
\begin{center}
\includegraphics[scale=0.5]{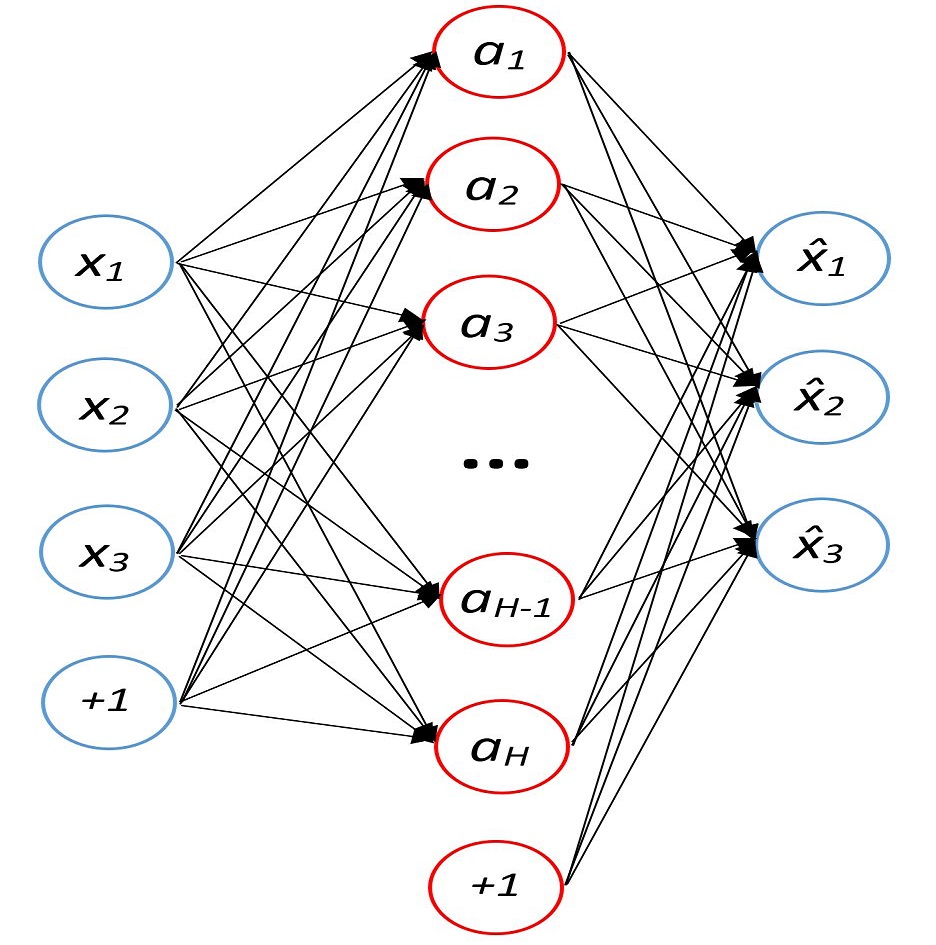}
\end{center}
\caption{Sparse Autoencoder}
\label{sparse autoencoder}
\end{figure}

Autoencoders are unsupervised artificial neural networks that learn to copy their inputs to their outputs under some constraints. The constraints are added at the hidden layer. They force the autoencoder to learn an efficient representation of the input data. Sparsity is such a constraint. In sparse autoencoders \cite{ng2011sparse}, the number of neurons in the hidden layer is usually greater than the number of inputs as shown in Figure \ref{sparse autoencoder}. Sparsity is added by forcing the autoencoder to reduce the number of active neurons in the hidden layer. A neuron is considered to be active if its output is close to 1 and inactive if its output is close to 0. For example, one can constrain the autoencoder to have on average 1\% of significantly active neurons in the hidden layer. To do so, first, the average activation of each neurons in the hidden layer is computed. Then neurons that have on average an activation greater than the targeted activation are penalized by adding a sparsity loss term to the cost function. The contribution of the sparsity loss term to the total cost function is controlled through a sparsity weight parameter.

The difference between the input and the output is called the reconstruction error. Let $\hat{x}=(\hat{x}_{1},\hat{x}_{2},…,\hat{x}_{n})$ be the output of the autoencoder when fed with a specific input $x=(x_{1},x_{2},…,x_{n})$. Then the reconstruction error $RE$ is given by:
\begin{center}
  $RE = \sum (\hat{x}_{i} - x_{i})^{2}$  
\end{center}

During the training phase, the parameters of an autoencoder are optimized in order to minimize the reconstruction error for a particular dataset. Once trained, if the autoencoder is fed with data that are similar to the data used during the training phase, the reconstruction error will be small. On the contrary, the reconstruction error will be large when test data are different from the data used during the training phase.

\subsection{Detection Threshold Determination} \label{detection threshold}
The reconstruction error $RE$ can be used as a measure of the outlierness of new samples. To use an autoencoder as an anomaly detector, a detection threshold has to be fixed. The detection threshold is the value of the reconstruction error above which an instance is considered as being anomalous.

The detection threshold is determined using a validation dataset that is different from the one used for training. The detection threshold $thr$ is calculated as follows:
\begin{center}
    $thr = \mu (RE_{Val}) + \sigma (RE_{Val})$
\end{center}

where $\mu (RE_{Val})$ is the mean reconstruction error over the validation set and $\sigma (RE_{Val})$ the corresponding standard deviation. Extreme outliers are removed before calculating the detection threshold. Extreme outliers are any validation sample for which the reconstruction error is greater than two times the total number of features ($2n$). Indeed, we can prove that if the reconstruction error is greater than twice the total number of features then the autoencoder is performing worse than random guessing. 
\begin{proof}
Let $X_{i}$ be a random variable that model the distribution of the $i^{th}$ feature of the input data.  As the data are normalized before being fed to the autoencoder, $X_{i}$ follows a normal distribution centered to 0 and with unit variance $X_{i} \sim N(0, 1)$. Let $\hat{X}_{i}$ be a random variable that model the $i^{th}$ feature of the output vector. Let us consider that we randomly guess the value of each feature of the output vector from a normal distribution centered to 0 and with unit variance, that is, $\hat{X}_{i} \sim N(0, 1)$. Then the error on each individual feature is given by:
\begin{center}
    $E[(X_{i} - \hat{X}_{i})^{2}] = Var(X_{i} - \hat{X}_{i}) + E[X_{i} - \hat{X}_{i}]^{2} = Var(X_{i}) + Var(\hat{X}_{i}) + Cov(X_{i}, \hat{X}_{i}) + 0 = 2$
\end{center}
where $E, Var$ and $Cov$ stand for the expectation, the variance and the covariance respectively. As the overall reconstruction error is equal to the sum of the error on each feature, we end up with a reconstruction error equal to $2n$. 
\end{proof}
In other words, if we randomly guess the value of each feature of the output vector from a normal distribution $N(0, 1)$, then on average the reconstruction error will be equal to $2n$.

\section{Proposed Model} \label{Proposed Model}

\subsection{Overview}

\begin{figure}[]
\begin{center}
\includegraphics[scale=0.51]{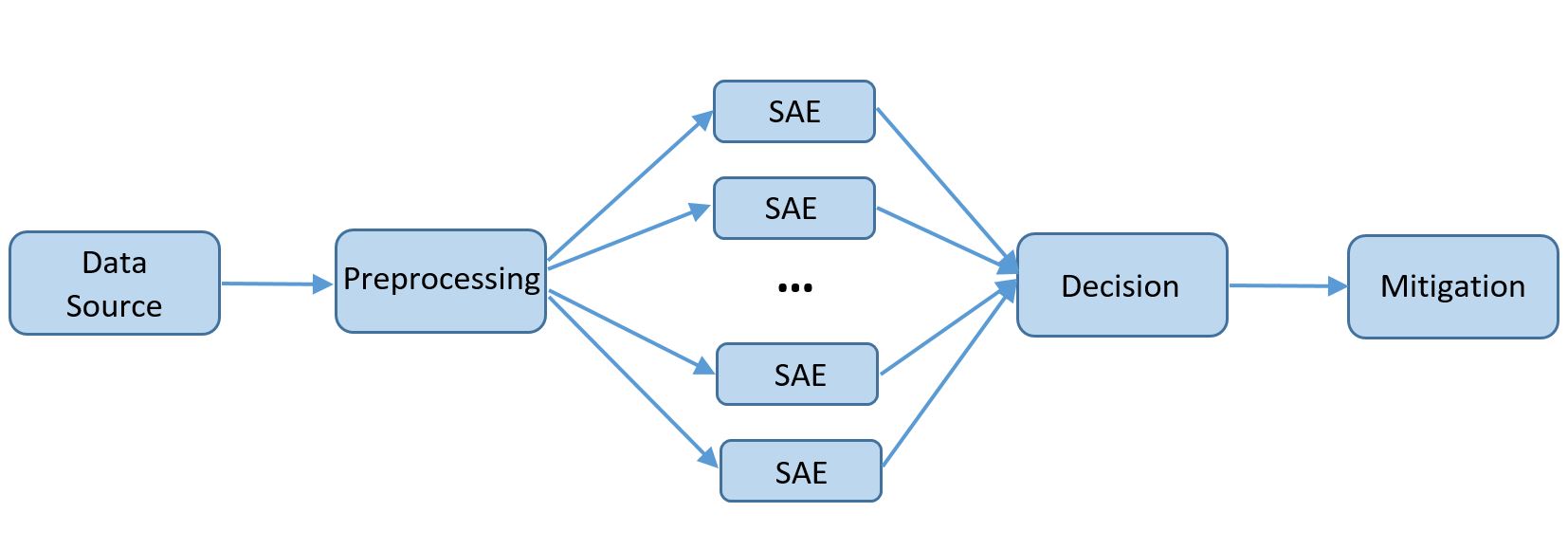}
\end{center}
\caption{Proposed anomalous communications detection architecture using a set of sparse autoencoders (SAE)}
\label{proposed architecture}
\end{figure}

We propose to detect anomalous communications in IoT networks using a set of sparse autoencoders. Figure \ref{proposed architecture} shows the proposed architecture. First, network communication data are preprocessed to extract useful features (descibed in Section \ref{features description}). The preprocessing step also include features normalization. The normalized data is then fed to multiple sparse autoencoders. Indeed, for each IoT device type present in the network, one different sparse autoencoder has been trained to learn its legitimate communication profile. Next, a decision module takes the outputs of all the sparse autoencoders and determine whether a communication is anomalous or not. Finally, the role of the mitigation module is to block communications detected as being anomalous.

\subsection{Decision Module} \label{final decision}

One different sparse autoencoder is trained to learn the profile of the legitimate communications for each IoT device type present in the network. By IoT device type we refer to a device model from a specific manufacturer. We assume that during the testing phase (or actual deployment phase) it is not possible to know what device is generating the ongoing network communications. Therefore, the data describing each communication are fed to all the trained sparse autoencoders. A communication is considered as being anomalous only if all autoencoders consider so. Formally, let us consider a total of $T$ different IoT device types. Hence, the communication data are fed to $T$ different sparse autoencoders. The reconstruction errors on the output of each autoencoder are calculated and compared to their respective detection thresholds to determine whether the communication is anomalous or not. Let $\{A_{1}(x), A_{2}(x),..., A_{T}(x)\}$ be the set of decisions obtained for the $T$ sparse autoencoders when fed with input $x$, with $A_{i}:\mathbb{R}^{n} \mapsto \{0, 1\}$ the decision of the sparse autoencoder trained to learn the legitimate network communication profile of the $i^{th}$ device type. It takes as input a network communication $x$ from the feature space $\mathbb{R}^{n}$ and output 0 if the communication is legitimate and 1 if it is anomalous. Note that we assume that for each IoT device type present in the network, there exists an autoencoder that has been trained to learn its legitimate communication profile. Let $x$ be the network communication being tested. It is fed to all the trained sparse autoencoders. The final decision $anomaly$ is given by:
\begin{center}
    $anomaly = \bigcap_{i=1}^{T} A_{i}(x)$
\end{center}
where $anomaly$ is equal to 1 if the communication is anomalous. Note that the communication is considered to be legitimate if at least one autoencoder considers so. The advantage of doing so is that we no longer need to know the type of IoT device that generated the network communication beforehand. Indeed, depending on where the detection system is located it can be difficult to find out the type of IoT device that generated the traffic. For example, when the devices communicate through a NAT proxy and the detection system is located outside the local network.

\subsection{Features Description} \label{features description}

Network traffic data are preprocessed in order to extract bidirectional TCP flows identified by their source and destination IP addresses and ports. A timeout is used to split long TCP connections into multiple bidirectional flows. The features describing the bidirectional TCP flows are statistics on the size of the first N packets sent and received, along with the corresponding inter-arrival times (IAT). The features are described in Table \ref{features}. 

\begin{table}[]
\centering
\caption{Features used to describe bidirectional TCP flows}
\label{features}
\begin{tabular}{|c|}
\hline
\textbf{Features}                                                                                                           \\ \hline
\begin{tabular}[c]{@{}c@{}}Mean, Median, Min, Max, Standard deviation and Count\\ of the size of the first N packets sent\end{tabular}     \\ \hline
\begin{tabular}[c]{@{}c@{}}Mean, Median, Min, Max, Standard deviation and Count\\ of the size of the first N packets received\end{tabular} \\ \hline
\begin{tabular}[c]{@{}c@{}}Mean and Standard deviation\\ of the IAT between the first N packets sent\end{tabular}                          \\ \hline
\begin{tabular}[c]{@{}c@{}}Mean and Standard deviation\\ of the IAT between the first N packets received\end{tabular}                      \\ \hline
\end{tabular}
\end{table}

Count stands for the number of packets among the first N packets that have a non-zero size. In other words, it corresponds to the actual number of packets sent or received. It can be less than N because of the timeout used to split long TCP connections. For example, if the total number of packets sent for the duration of the timeout is equal to some value less than N, then count is equal to that value. Otherwise, it is equal to N. For the IAT to exist, N should be equal or greater than 2. Note that if a communication contains only one packet sent or received, the statistics that cannot be calculated, such as, the mean and standard deviation of the IAT between packets, are all set to 0.

The features used to describe the network traffic are application agnostic. Any application can run on top of the TCP protocol. Note that the features can be extracted even if the network communication is encrypted. The size of packets and IAT have been successfully used in other IoT related works \cite{IoTDevicesRecognition}\cite{nbiotMeidan}\cite{DoshiDDOS}. 
Our study focuses only on TCP protocols as all the devices used for experiments use HTTP/HTTPS for communications. This makes sense as most of the network communications generated by IoT malware use TCP \cite{symantecISTR}. However, the proposed method can be easily extended to UDP communications.

\section{Dataset Description} \label{dataset description}

To train and test the model, one need three different datasets: a training set, a validation set and a test set. The training set is used to optimize the parameters also called the weights of the neural network. The validation set is used to fine tune the hyperparameters of the model, namely, the learning rate, the number of epochs used for training (early stopping) and the decision threshold (described in Section \ref{detection threshold}). Both the training and validation sets only contain legitimate network traffic data as they are used to learn the normal communication profile. The test set is used to assess the performance of the developed model and it contains both legitimate and malicious network traffic data. 

A small smart home network is set up to generate legitimate network communications. The experimental smart home consists of four IoT devices: a Nest security camera, a D-Link motion sensor, a TP-Link smart bulb and a TP-Link smart plug. 
Network traffic has been collected for a total of 7 days. Table \ref{legitimate flow} shows the total number of bidirectional TCP flows extracted for each device. 


\begin{table}[]
\centering
\caption{Total number of bidirectional flows per device}
\label{legitimate flow}
\begin{tabular}{c|c|}
\cline{2-2}
                                           & Bidirectional flows \\ \hline
\multicolumn{1}{|c|}{D-Link Motion Sensor} & 1074                \\ \hline
\multicolumn{1}{|c|}{Nest Security Camera} & 1055                \\ \hline
\multicolumn{1}{|c|}{TP-Link Smart Bulb}   & 1040                \\ \hline
\multicolumn{1}{|c|}{TP-Link Smart Plug}   & 858                 \\ \hline
\multicolumn{1}{|c|}{\textbf{Total}}       & \textbf{4027}       \\ \hline
\end{tabular}
\end{table}

Malicious bidirectional flows used in the test set are obtained from IoTPOT \cite{iotpot}, an IoT honeypot designed to get infected by IoT malware. A total of 46,796 bidirectional TCP flows are extracted from one day of network traffic data. The collected network traffic data represents network traffic generated by IoT botnets. Note that the number of malicious bidirectional flows is much larger than the number of legitimate ones. This is not an issue as the malicious flows are only used during the testing phase to assess the attack detection rate.

\section{Evaluation} \label{experimental results}

\subsection{Experimental Setup}

As described in Section \ref{final decision}, one sparse autoencoder per device is trained to learn the legitimate communication profile. The architecture of each sparse autoencoder for our experiment is as follows: 
\begin{itemize}
\item The size of the input layer is 16 (equal to the number of features)
\item The size of the hidden layer is 32
\item The target sparsity is equal to 0.1
\item The sparsity weight is equal to 0.2
\end{itemize}

We train and test the model for different values of N (introduced in Section \ref{features description}) ranging from 2 to 10. For example, if N is equal to 5, the statistics described in Section \ref{features description}, are calculated over the first 5 packets sent and received. To get the best out of the limited number of legitimate communication samples, we perform 5-fold cross-validation. That is, the set of collected legitimate communications is split into 5 folds. The sparse autoencoders are trained using 4 folds (that actually consists of the training set and the validation set as described in Section \ref{dataset description}) and the remaining fold is used as the test set. The process is repeated 5 times, with each of the 5 folds used exactly once as the test set.

\subsection{Achieved Performance}

The true positive rate (also called recall or attack detection rate) and the false positive rate are used to assess the performance of the model. Let $TP$, $TN$, $FP$ and $FN$ be the number of true positive, true negative, false positive and false negative respectively. The true positive rate ($TPR$) and the false positive rate ($FPR$) are given by the following formula:
\begin{center}
$TPR = \frac{TP}{TP + FN}$
\end{center}
\begin{center}
$FPR = \frac{FP}{FP + TN}$
\end{center}

\begin{figure}[]
\begin{center}
\includegraphics[scale=0.45]{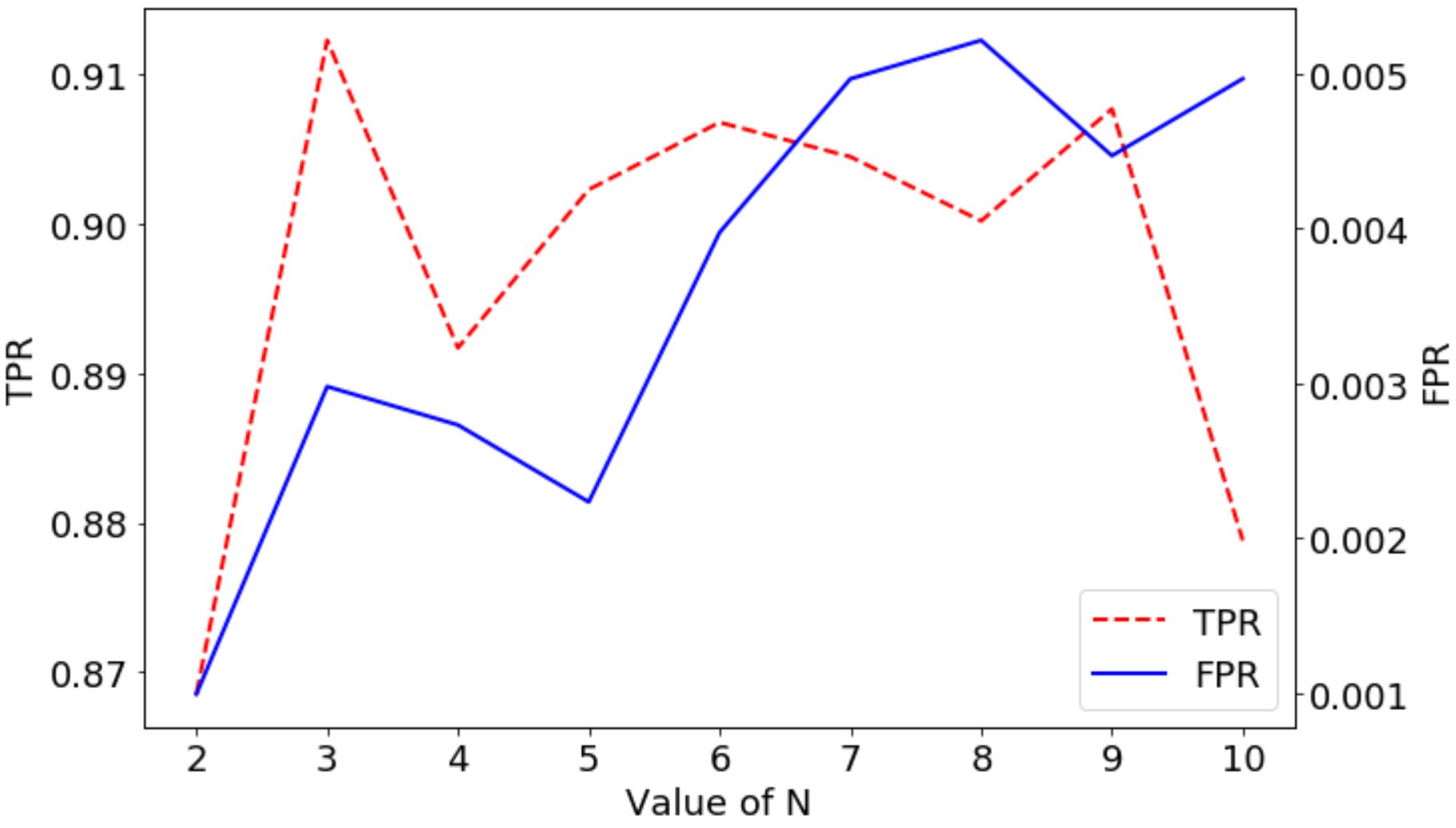}
\end{center}
\caption{Achieved $TPR$ and $FPR$ for different values of N}
\label{TPR_FPR}
\end{figure}

To assess the performance of the overall model, the test data is fed to all the sparse autoencoders and the final decision is obtained following the method described in Section \ref{final decision}. Figure \ref{TPR_FPR} shows the achieved $TPR$ and $FPR$ for different values of N. $TPR$ oscillates between 86.9\% and 91.2\%. The $TPR$ reaches the maximum for N equal to 3. That is, looking at more packets than the first 3 packets sent and received no longer improve the $TPR$. As for the $FPR$, it ranges from 0.1\% to 0.5\%, depending on the value of N. $FPR$ tends to increase as N goes up. This can be explained by the presence of legitimate samples in the test set that behave more like outliers above a certain value of N. For example, if the size of the first 6 packets of a communication are stable and in accordance with what was learnt by the model to be legitimate then it will be correctly classified as being legitimate by any model trained for a value of N less or equal to 6. But if that same communication has the size of its 7th sent packet unduly small or large, then some features such as the mean size of the first 7 packets sent will be greatly affected. As a consequence, models trained for a value of N greater than or equal to 7 will misclassify this same communication as being anomalous. Hence, once a legitimate communication has been misclassified as being anomalous by a model trained for a specific value of N, it is very unlikely for that same communication to be correctly classified by a model trained for a greater value of N. 
This is why $FPR$ tends to increase with N.

\section{Related Work} \label{Related Work}

Few works about machine learning for IoT network intrusion detection exist. In \cite{nbiotMeidan}, stacked vanilla autoencoders are used to detect compromised devices in a network. 
However, the work does not aim to separate malicious communications from legitimate ones, but only to detect whether or not a device is infected. The work presented in \cite{diotNguyen} takes advantage of the temporal periodicity of traffic generated by IoT devices to detect compromised devices. They use Gated Recurrent Units with a detection threshold. In \cite{LuoDistributedADinWSN}, the authors use an autoencoder to detect anomalies in the data read by the nodes of a Wireless sensor network (WSN).
Other works leverage supervised machine learning. In \cite{moustafaEnsembleIDSstatisticalflowfeatures_MQTT_DNS_HTTP}, a set of features is defined to describe MQTT, DNS and HTTP protocols. An AdaBoost ensemble learning algorithm, composed of three different machine learning models, is developed to detect attacks. In \cite{DoshiDDOS}, supervised machine learning is used to perform DDoS attack detection in consumer IoT networks. 
Other works focus on leveraging machine learning to fingerprint IoT devices in a network \cite{IoTDevicesRecognition}\cite{MiettinenSENTINEL}\cite{meidanUnauthorized}\cite{MeidanProfillIoT}. The purpose of IoT device type fingerprinting is to detect devices that are considered to be vulnerable in order to deny access to the network.

\section{Conclusion and Future Works} \label{Conclusion and Future Works}

This paper introduced a method to detect anomalous communications in IoT networks using a set of sparse autoencoders. One different sparse autoencoder is trained to learn the legitimate communication profile of each IoT device type present in the network. The features used are statistics on the size of the first N packets sent and received, along with statistics on the inter-arrival times between the packets. We set up an experimental smart home network to assess the performance of our model. Depending on the value of N, our model achieved attack detection rates ranging from 86.9\% to 91.2\% and false positive rates ranging from 0.1\% to 0.5\%. 
Once a communication is detected as being anomalous it can be dropped without having to totally interrupt the service provided by the device.

For future works, our model should be tested with a network that contains a larger number of IoT devices. One should also consider the case of a general purpose network that contains not only IoT devices but also personal computers or smartphones. Adversarial machine learning can also be leveraged to assess the robustness of our proposed model. We are also planning to integrate the developed anomaly detection model in a Software Defined Networking (SDN) environment in order to provide centralized monitoring and management of multiple IoT networks.

\bibliographystyle{IEEEtran}

\bibliography{bib}

\end{document}